\documentclass[manuscript]{aastex61}

\shorttitle{The first study of AH Mic contact system}
\usepackage{lineno}
\usepackage{amsmath}
\usepackage{graphicx}
\usepackage{xcolor}
\usepackage{longtable}
\usepackage{times}
\usepackage{tablefootnote}
\usepackage{appendix}
\usepackage{lineno}

\begin{document}

\title{The First Photometric Study of AH Mic Contact Binary System}

\correspondingauthor{Atila Poro}
\email{astronomydep@raderonlab.ca and poroatila@gmail.com}

\author{A. Poro}
\affil{Astronomy Department of the Raderon Lab., BC, Burnaby, Canada}

\author{M.G. Blackford}
\affil{Variable Stars South (VSS), Congarinni Observatory, Congarinni, NSW, 2447, Australia}

\author{S. Ranjbar Salehian}
\affil{Binary Systems of South and North (BSN) Project, Contact Systems Department, Iran}

\author{E. Jahangiri}
\affil{Binary Systems of South and North (BSN) Project, Contact Systems Department, Iran}

\author{M. Samiei Dastjerdi}
\affil{Binary Systems of South and North (BSN) Project, Contact Systems Department, Iran}

\author{M. Gozarandi}
\affil{Binary Systems of South and North (BSN) Project, Contact Systems Department, Iran}

\author{R. Karimi}
\affil{Binary Systems of South and North (BSN) Project, Contact Systems Department, Iran}
\affil{Department of Physics, Kharazmi University, Tehran, Iran}

\author{T. Madayen}
\affil{Astronomy and Astrophysics Department, University of Toronto, Toronto, Canada}

\author{E. Bakhshi}
\affil{Binary Systems of South and North (BSN) Project, Contact Systems Department, Global Online Project, Iran}
\affil{Department of Physics, Institute for Advanced Studies in Basic Sciences (IASBS), Zanjan, Iran}

\author{F. Hedayati}
\affil{Binary Systems of South and North (BSN) Project, Contact Systems Department, Iran}

\begin{abstract}
The first multi-color light curve analysis of the AH Mic binary system is presented. This system has very few past observations from the southern hemisphere. We extracted the minima times from the light curves based on the Markov Chain Monte Carlo (MCMC) approach and obtained a new ephemeris. To provide modern photometric light curve solutions, we used the Physics of Eclipsing Binaries (Phoebe) software package and the MCMC approach. Light curve solutions yielded a system temperature ratio of $0.950$, and we assumed a cold star-spot for the hotter star based on the O'Connell effect. This analysis reveals that AH Mic is a W-subtype W UMa contact system with a fill-out factor of 21.3\% and a mass ratio of 2.32. The absolute physical parameters of the components are estimated by using the Gaia Early Data Release 3 (EDR3) parallax method to be $M_h(M_\odot)=0.702(26)$, $M_c(M_\odot)=1.629(104)$, $R_h(R_\odot)=0.852(21)$, $R_c(R_\odot)=1.240(28)$, $L_h(L_\odot)=0.618(3)$ and $L_c(L_\odot)=1.067(7)$. The orbital angular momentum of the AH Mic binary system was found to be $51.866(35)$. The components' positions of this system are plotted in the Hertzsprung-Russell (H-R) diagram.
\end{abstract}


\section{Introduction}
Each component of a contact binary system has its own Roche lobe, with the inner and outer critical surfaces sharing a common envelope. These stars are generally late type F, G, or K type stars. The majority of contact systems are solar-type dwarfs with active stellar populations, as evidenced by their asymmetric light curves and spectral line profiles.

Contact binaries typically have orbital periods of less than one day, and the \cite{2020MNRAS.497.3493Z} investigation found that the lower limit of orbital period is 0.15 days. The orbital period is frequently varied over time, due to mass transfer from one component to the other, as well as mass and angular momentum loss from the binary system. Although, according to \cite{2021ApJS..254...10L} study, systems with orbital periods of more than 0.5 days and temperatures greater than 7000 K probably have radiative envelopes and should not be classified as W UMa-type contact binaries.

Contact binaries are debatable in terms of mass, and the lower limit of mass ratio for these systems is presently under investigation (e.g. \citealt{2015AJ....150...69Y}; \citealt{2007MNRAS.377.1635A}; \citealt{2006MNRAS.369.2001L}). Additionally, since many investigations lack a spectroscopic method, determining the mass ratio is a challenging issue.

W Ursae Majoris (W UMa)-type contact systems (EW) are classified into two categories, W-type and A-type, based on their properties. The mass ratio in A-subtype systems is often less than 0.5, showing weak or moderate activity. In W-subtype systems the less massive component is hotter and the period changes continuously over time (\citealt{1970VA.....12..217B}). In comparison with the W-subtypes, A-subtype systems
are earlier spectral types with higher mass and luminosity.

The purpose of this study is to present the first light curve analysis of the AH Mic binary system and determine its properties. AH Mic is a binary system from the southern hemisphere with a range of apparent magnitude of $V=12.85-13.45$ and an orbital period of $0.3243344$ days (\citealt{2006SASS...25...47W}). In all the available catalogs, this binary system is classified as a EW-type system. This paper is arranged as follows: Section 2 contains information about the multi-color photometric observations from an Australian observatory as well as a data reduction technique. Section 3 explains how to extract our times of minima, collect minima collection times from literature, and compute a new ephemeris. The Phoebe Python code and the MCMC approach were used to study the light curves of the AH Mic system, and the absolute parameters were obtained in section 4. Section 5 has a summary and conclusion.

\vspace{1.5cm}
\section{Observation and data reduction}
\begin{figure}
\begin{center}
\includegraphics[width=13.5cm,height=9cm]{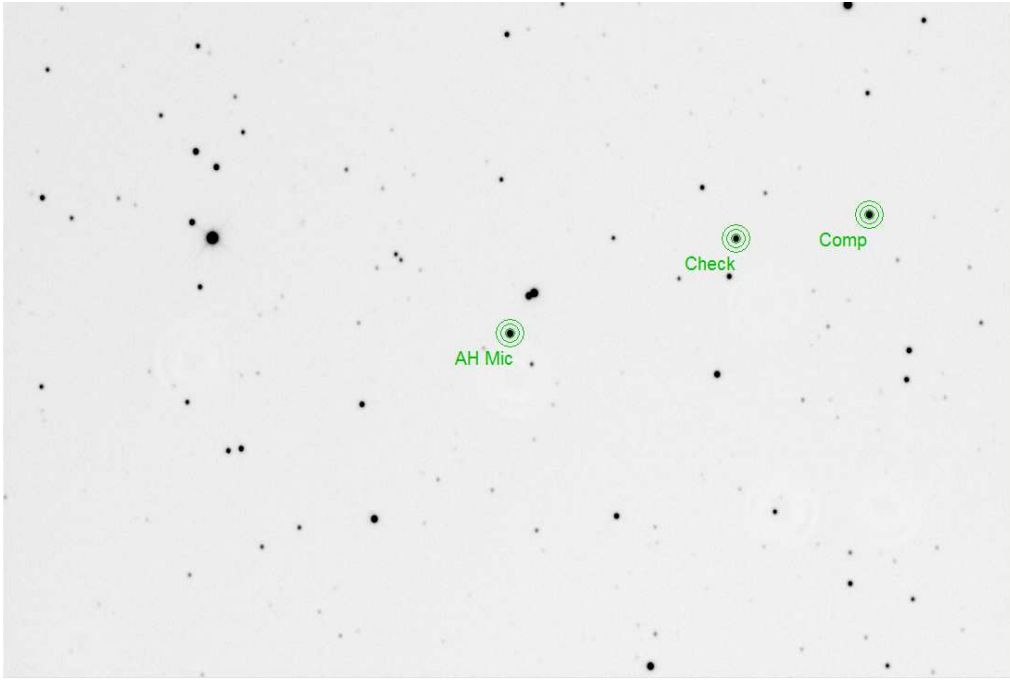}
    \caption{AH Mic, comparison star, and check star field-of-view. The circles are solely used to show the position of each star, which is indicated by bigger rings.}
\label{Fig1}
\end{center}
\end{figure}

AH Mic (RA. $21$ $04$ $57.8280$, Dec. $-40$ $33$ $06.084$ (J2000)) was observed for six nights in August and September 2020 with a GSO 14-inch f/8 Ritchey Chretien telescope at the Congarinni Observatory in Australia ({$152^\circ$} $52'$ East and {$30^\circ$} $44'$ South). The data was taken using a SBIG STT3200-ME CCD camera ($2184\times1510$ pixels, 6.8 micron square) with $2\times2$ binning and a CCD temperature of $-15^\circ$C. These observations were made using the Astrodon Johnson-Cousins $BVRI$ standard filters. The exposure time of these observations for $BI$ filters is 120 seconds and for $VR$ filters is 60 seconds.
GSC 7969-723 was selected as a comparison star and GSC 7969-1170 was chosen as a check star with an appropriate apparent magnitude in comparison to AH Mic.
The comparison star was found at R.A. 21h 04m 49.91976s (J2000), Dec. $-40^\circ$ $39'$ $42.8868"$ (J2000) with a $V=12.831$ magnitude, while the check star was located at R.A. 21h 04m 50.96544s (J2000), Dec. $-40^\circ$ $37'$ $18.0120"$ (J2000) with a $V=13.285$ magnitude, based on the	AAVSO Photometric All Sky Survey DR9 (APASS9) catalog. Figure \ref{Fig1} shows the observed field-of-view for AH Mic with the comparison and the check stars.

During the observations, a total of 1417 images were acquired. The CCD image processing was done with MaxIm DL software, which included dark, bias, and flat-field corrections (\citealt{2000IAPPP..79....2G}). The airmass decreases the flux of stars when light passes through the Earth's atmosphere. We estimated the airmass based on observatory coordinates and the position of the star in the sky during the observations (\citealt{1962aste.book.....H}). Finally, we used the AstroImageJ (AIJ) software to normalize all of the ground-based data (\citealt{2017AJ....153...77C}), and apply airmass\footnote{This study's data (after data reduction) is available in four filters as a supplement file.}.

\vspace{1.5cm}
\section{New ephemeris calculation}
We extracted the nine times of minima in the light curves which four of them are the primary minima. We used models based on the Gaussian and Cauchy distributions to find the times of minima and the MCMC sampling methods to estimate the uncertainty of the values (\citealt{2021AstL...47..402P}). The Python code for this extraction is implemented using the PyMC3 package (\citealt{2016ascl.soft10016S}). All times of minimum are expressed in the Barycentric Julian Date in the Barycentric Dynamical Time ($BJD_{TDB}$), and table \ref{tab1} shows all CCD times of minima with their uncertainties of AH Mic system.

O-C variations are the differences between observed mid-eclipse times (O) and their calculated values (C) based on a reference ephemeris. An observable trend in O-C is caused by various factors and effects. Due to the small number of minimum times and the short interval of observation (2013–2020), only a linear fit can be considered for the AH Mic system. We fitted a line on all times of minima to calculate a new ephemeris based on the MCMC approach using the emcee package in Python (\citealt{2013PASP..125..306F}) as shown in Figure 2.

We determined a new ephemeris for primary minimum as follows:

\begin{equation}\label{eq1}
BJD_{TDB}(MinI.)=(2456523.00559\pm0.00068)+(0.32433351\pm0.00000009)\times E.
\end{equation}

\begin{table*}
\caption{Available CCD times of minima for AH Mic.}
\centering
\begin{center}
\footnotesize
\begin{tabular}{c c c c c}
\hline
\hline
$Min.(BJD_{TDB})$	& Error	& Epoch	& O-C	& Reference\\
\hline
$2456523.00509$	& $0.00040$	& $0$	& $0$	& \cite{2014IBVS.6093....1D} \\
$2457154.80792$	& $0.00027$	& $1948$	& $-0.0006$	& \cite{2019MNRAS.486.1907J} \\
$2459108.91600$	& $0.00017$	& $7973$	& $-0.0073$	& This study \\
$2459089.94354$	& $0.00010$	& $7914.5$	& $-0.0062$	& This study \\
$2459090.10578$	& $0.00007$	& $7915$	& $-0.0061$	& This study \\
$2459090.91664$	& $0.00008$	& $7917.5$	& $-0.0061$	& This study \\
$2459091.07740$	& $0.00013$	& $7918$	& $-0.0075$	& This study \\
$2459091.24157$	& $0.00016$	& $7918.5$	& $-0.0055$	& This study \\
$2459092.05120$	& $0.00050$	& $7921$	& $-0.0067$	& This study \\
$2459092.21189$	& $0.00092$	& $7921.5$	& $-0.0081$	& This study \\
$2459109.07933$	& $0.00020$	& $7973.5$	& $-0.0061$	& This study \\
\hline
\hline
\end{tabular}
\end{center}
\label{tab1}
\end{table*}

\begin{figure}
\begin{center}
\includegraphics[width=13.5cm,height=9cm]{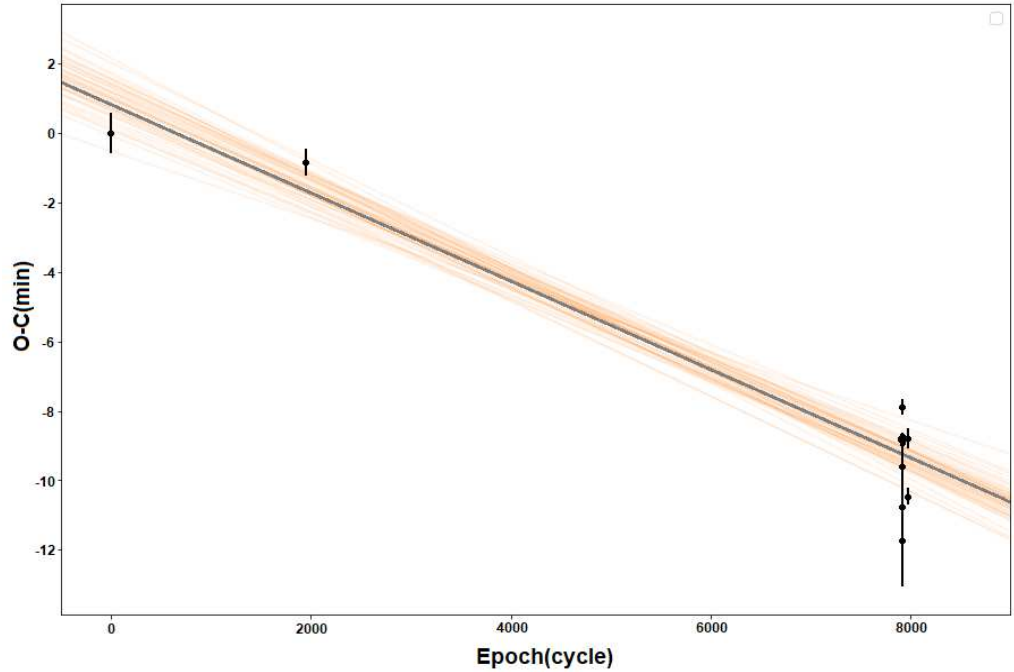}
    \caption{The O-C diagram of AH Mic system.}
\label{Fig2}
\end{center}
\end{figure}

\vspace{1.5cm}
\section{Photometric Solutions}
\subsection{Light curve analysis}
We utilized the latest Python version of the Phoebe software package (2.3.58) to model the light curves of the AH Mic binary system. We employed the Phoebe along with the emcee package (\citealt{2013PASP..125..306F}), with the MCMC approach for more precision in modeling, output results, and uncertainties. We applied 32 walkers and 1000 iterations to each walker in the MCMC approach. The Phoebe and MCMC approach then searched for the temperature ratio, orbital inclination, and mass ratio.
The bolometric albedo and gravity-darkening coefficients were assumed to be $g_h=g_c=0.32$ (\citealt{1967ZA.....65...89L}) and $A_h=A_c=0.5$ (\citealt{1969AcA....19..245R}). The stellar atmosphere was modeled using the \cite{2004A&A...419..725C} method, and the limb-darkening coefficients were employed as a free parameter in the Phoebe. 
Different methods were used to determine the effective temperature for the hotter component as a fixed parameter. (1) The temperature of the Gaia DR2\footnote{https://gea.esac.esa.int/archive/}  for AH Mic system is $5381_{\rm-(51)}^{+(74)}$ K; (2) The result of the $B-V$ index obtained from this work's light curves to be $0.73\pm0.04$ after calibration (\citealt{2000A&A...357..367H}) and as a result we found the effective temperature $5470_{\rm-(119)}^{+(111)}$ K according to \cite{1996ApJ...469..355F} study; (3) The effective temperature from the result of \cite{2022MNRAS.510.5315P} analysis of the relationship between orbital period and the primary (hotter) temperature for contact systems is $5681_{\rm-(80)}^{+(81)}$ K. We chose Gaia DR2's temperature as a fixed parameter for the hotter star. In the continuation, we performed the initial light curve analysis. Then, based on the following equations from \cite{2018PASA...35....8K} study, we calculated the temperatures and performed the final photometric light curve solutions:

\begin{equation}\label{eq2}
T_1=T_m+\frac{c\Delta T}{c+1}
\end{equation}

\begin{equation}\label{eq3}
T_2=T_1-\Delta T
\end{equation}

where $\Delta T$ is the temperature difference between the two stars, and $c$ is $l_2$ divided by $l_1$.

The $BVRI$ light curves of the AH Mic system are asymmetric in the maxima. This shows that the O'Connell effect (\citealt{1951PRCO....2...85O}) is present in this system, as $Max II$ is brighter than $Max I$. We employed a stellar spot model during the light curve solution in all used filters, and we found that assuming a cool star-spot model on the hotter component leads to the appropriate solutions for all $BVRI$ light curves (Table \ref{tab2}). Table \ref{tab3} shows the characteristic properties of the AH Mic light curves, with the difference in maxima in each filter displayed in the first row. Therefore, the $B$ filter had the highest difference between the levels of maximum light and the depths of the minima, while the $R$ filter had the least. This is expected because of the induced changes in the magnetic activity.

\begin{figure}
\begin{center}
\includegraphics[width=\columnwidth]{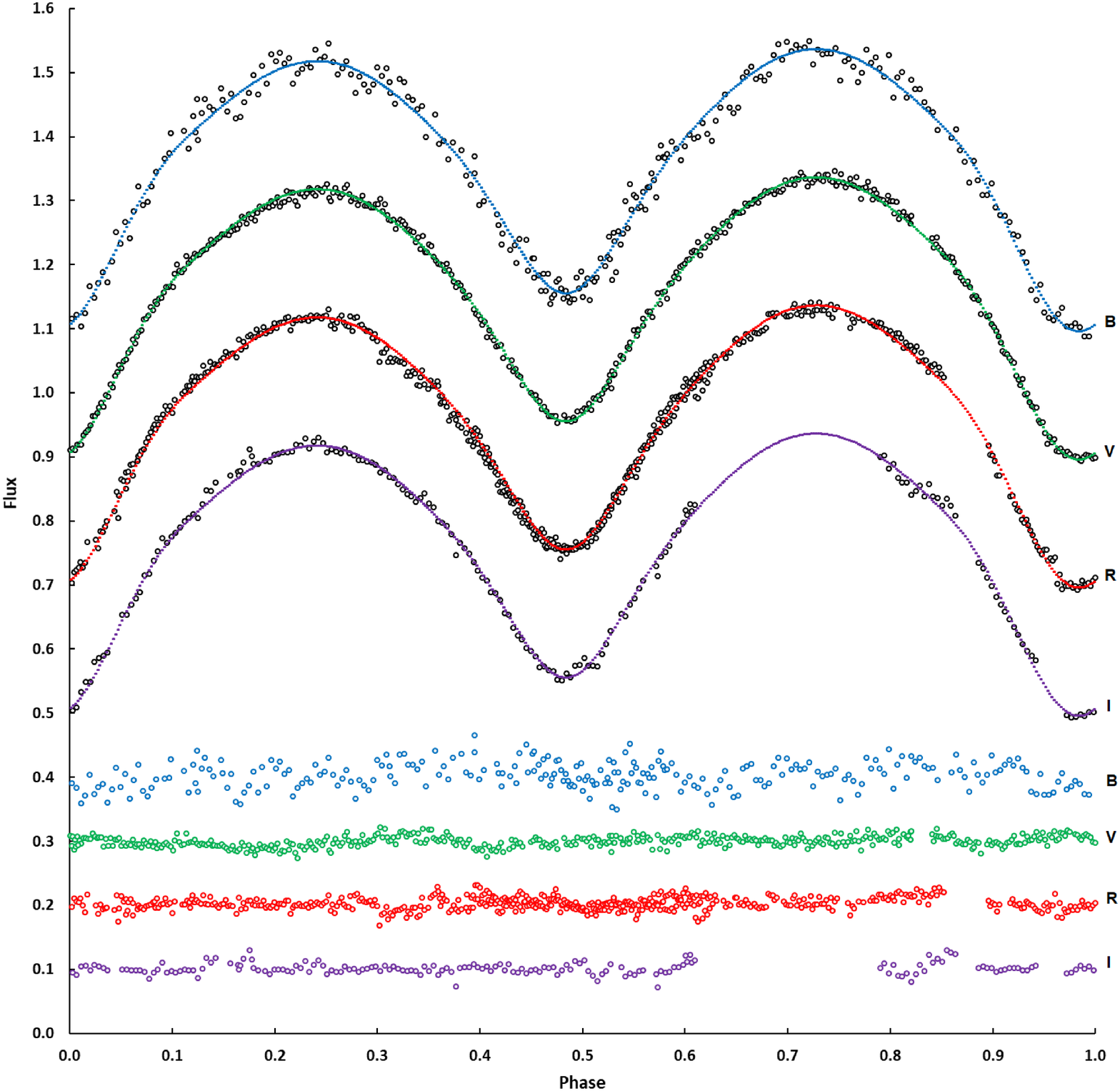}
    \caption{The observed light curves of AH Mic (black dots), and synthetic light curves obtained from light curve solutions and residuals are plotted, with respect to orbital phase, shifted arbitrarily in the relative flux.}
\label{Fig3}
\end{center}
\end{figure}

\begin{figure}
\begin{center}
\includegraphics[width=\columnwidth]{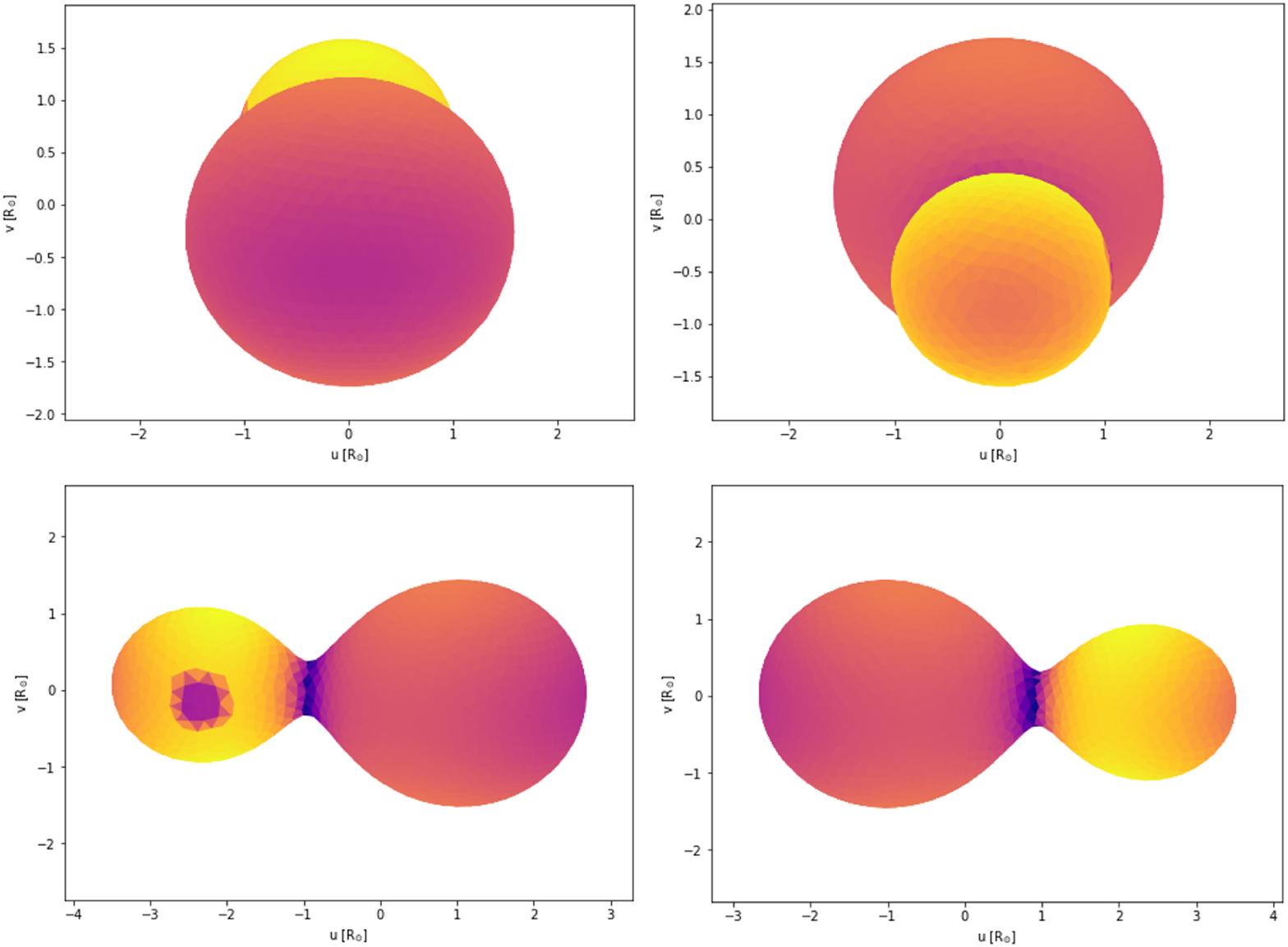}
    \caption{The positions of the components of AH Mic.}
\label{Fig4}
\end{center}
\end{figure}

Table \ref{tab2} shows the results of the light curve analysis, whereas Figure \ref{Fig3} displays the observational and theoretical light curves of our modeling. Figure \ref{Fig4} depicts the 3D view and geometrical structure of the AH Mic binary system.

\begin{table*}
\caption{Photometric solutions of AH Mic.}
\centering
\begin{center}
\footnotesize
\begin{tabular}{c c}
 \hline
 \hline
Parameter & Result\\
\hline
$T_{h}$ (K) & $5550_{\rm-(51)}^{+(74)}$\\
$T_{c}$ (K) & $5274_{\rm-(63)}^{+(40)}$\\
$q$ & $2.32_{\rm-(4)}^{+(8)}$\\
$\Omega_h=\Omega_c$ & $5.57_{\rm-(13)}^{+(48)}$\\
$i^{\circ}$ &	$75.91_{\rm-(35)}^{+(79)}$\\
$f$ & $0.213_{\rm-(12)}^{+(35)}$\\
$l_h/l_{tot}$ & $0.381(2)$\\
$l_c/l_{tot}$ & $0.619(2)$\\
$r_{h(mean)}$ & $0.324(2)$\\
$r_{c(mean)}$ & $0.470(2)$\\
Phase shift & $-0.005(1)$\\
\hline
$Colatitude(deg)$ & $90(2)$\\
$Longitude(deg)$ & $274(1)$\\
$Radius(deg)$ & $20(1)$\\
$T_{spot}/T_{star}$ & $0.91(2)$\\
\hline
\hline
\end{tabular}
\end{center}
\label{tab2}
\end{table*}

\begin{table*}
\caption{Characteristic parameters of the observational light curves in BVRI filters.}
\centering
\begin{center}
\footnotesize
\begin{tabular}{c c c c c}
 \hline
 \hline
Light curve	& $\Delta B^{(mag.)}$ & $\Delta V^{(mag.)}$ & $\Delta R^{(mag.)}$ & $\Delta I^{(mag.)}$\\
\hline
$Max I-Max II$ & 0.021 & 0.020 & 0.015 & \\
$Max I-Min II$ & -0.479 & -0.494 & -0.472 & -0.468\\
$Max I-Min I$ & -0.618 & -0.613 & -0.574 & -0.552\\
$Min I-Min II$ & 0.139 & 0.119 & 0.102 & 0.084\\
\hline
\hline
\end{tabular}
\end{center}
\label{tab3}
\end{table*}

\subsection{Absolute parameters}
The Gaia EDR3 parallax method can be used to estimate the absolute parameters when only photometric data is available. Several elements from observational and light curve analysis are used in the computations, despite the fact that the Gaia EDR3 parallax has high accuracy for this estimation. Recent investigations have made some corrections to the Gaia EDR3 parallax (e.g. \citealt{2021A&A...649A...4L}; \citealt{2021ApJ...911L..20R}). The range of these corrections appears to include the value of observational and light curve solution uncertainties (\citealt{2022MNRAS.510.5315P}).

Therefore, in order to properly estimate the absolute parameters of the stars of AH Mic binary system, we used the Gaia EDR3 parallax method. This is a well-known method, which is fully described in the \cite{2022MNRAS.510.5315P} study.

To perform the estimation, we first obtained the absolute magnitude ($M_v$) of the system using the maximum light of apparent magnitude $V_{max}=12.79(5)$ from our observation, the star's distance from Gaia EDR3 $d_{(pc)}=470.2(6.5)$, and the extinction coefficient $A_v=0.08$ from the \cite{2011ApJ...737..103S} study. Then, by using the value of $l_h/l_{tot}$ and $l_c/l_{tot}$ from the light curve solutions, we obtained the $M_v$ value of each of the component.
In addition, the bolometric magnitude $M_{bol}$ value of each star were obtained using $BC_h=-0.133$ and $BC_c=-0.199$. Furthermore the luminosity ($L$) values were estimated and the radius values of each star were obtained using the temperature of the stars from the light curve solutions. The semi-major axis of the system was also determined by utilizing $a=R/r_{(mean)}$ equation. According to Kepler's third law, the value of $M_h+M_c$ could be estimated using the semi-major axis and the orbital period of the system. Moreover, the mass of the hotter and the cooler stars were determined using the value of mass ratio.

To estimate the uncertainties using mass ratio and temperature from light curve analysis, we used the average value of uncertainties rather than the upper and lower ranges. The results of estimating the absolute parameters for the AH Mic system are shown in Table \ref{tab4}.

For comparison, we calculated absolute magnitude of system using the orbital period–luminosity calibration of \cite{2006MNRAS.368.1319R}:

\begin{equation}\label{eq4}
M_V=(-1.5)-12logP
\end{equation}

The result of relation 4 shows $M_V^{(mag.)}=4.37$, while in our estimation $M_V^{(mag.)}=4.349(10)$ which show relatively close values.

\begin{table*}
\caption{Estimation of absolute parameters of AH Mic binary system.}
\centering
\begin{center}
\footnotesize
\begin{tabular}{c c c c c c c}
 \hline
 \hline
Parameter & Hotter star & Cooler star\\
\hline
$M(M_\odot)$ & 0.702(26) & 1.629(104)\\
$R(R_\odot)$ & 0.852(21) & 1.240(28)\\
$L(L_\odot)$ & 0.618(3) & 1.067(7)\\
$M_{bol}(mag.)$ & 5.263(5) & 4.670(7)\\
$log(g)(cgs)$ & 4.423(5) & 4.463(8)\\
$a(R_\odot)$ & 2.634(48) & \\
\hline
\hline
\end{tabular}
\end{center}
\label{tab4}
\end{table*}

\vspace{1.5cm}
\section{Summery and conclusion}
Observations were made on the $BVRI$ filters for the AH Mic binary system in the Southern Hemisphere. The MCMC approach was used to extract the times of minima of our light curves, and the times of minima were also collected from previous observations. Accordingly, we presented a new ephemeris. The first light curve analysis of the AH Mic system was done in this work, which was performed using the latest version of the Phoebe Python code and the MCMC approach. To estimate the system's absolute parameters, we used the Gaia EDR3 parallax method, and the rest of the parameters were calculated using the observational parameters and the light curve solutions from this study.

We fixed the Gaia EDR3 temperature for the hotter component at first, and obtained the cooler component temperature from light curve solutions. This temperature is from the combination of both stars in the system. So, using Equation 2, we obtained the final temperature values of each star. The temperature difference between the two components is 276 K. According to Allen’s table (\citealt{2000asqu.book.....C}), these temperatures indicate that the spectral types of the hotter and cooler components are G5 and G8 respectively.

The mass ratio is an important parameter for contact binaries and it is determined through $q$-search when there is just photometric observations. The mass ratio in this study was determined using a method based on the MCMC approach. The reliable value of mass ratio through photometric light curve analysis is still under many investigations to approach the accuracy of this value using the spectroscopy method. However, investigations have been conducted, one of which is based on the Machine Learning methods using the Multi-Layer Perceptron (MLP) regression model and was carried out by the \cite{2022MNRAS.510.5315P} study. According to this method, we find the estimated range of mass ratio for AH Mic to be $0.408$ to $0.572$ with a mean of $0.485$, which includes our calculated mass ratio of $0.431$ from the light curve solutions. It's worth mentioning that the $q$ specified in the \cite{2022MNRAS.510.5315P} study is the ratio of the less massive star mass to the more massive star mass, hence $q$ is never more than unity.

The Orbital Angular Momentum ($J_0$) of the AH Mic system was calculated, and its location is displayed on the $logJ_0-logM$ diagram (Figure 5). The value of $logJ_0$ of AH Mic was obtained to be $51.866\pm0.035$. The diagram shows that this system is in a contact binary systems region.

\begin{figure}
\begin{center}
\includegraphics[width=13.5cm,height=9cm]{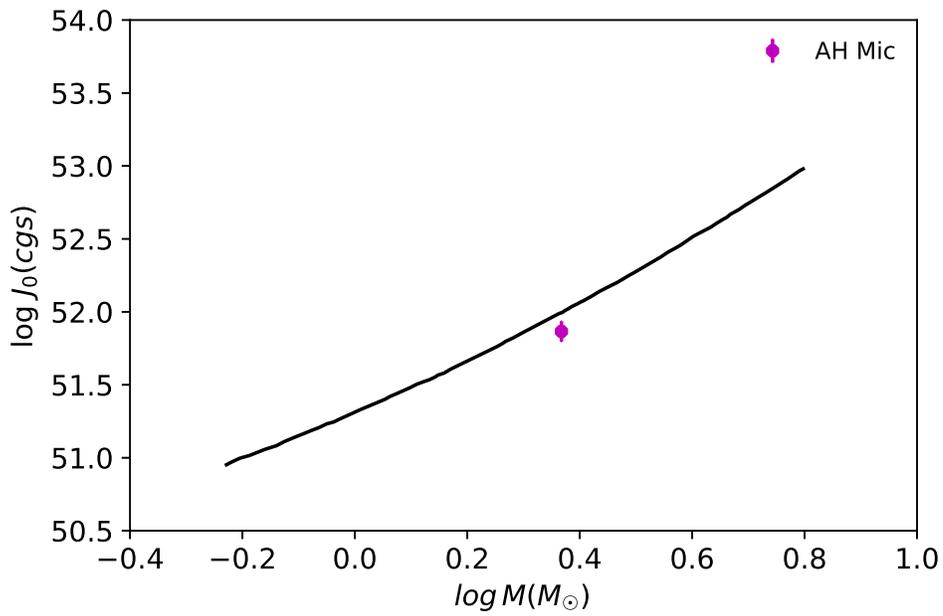}
    \caption{The location of AH Mic on the $logJ_0-logM$ diagram. Detached and contact binary systems are separated by a quadratic border line. The quadratic line is based on a study by \cite{2006MNRAS.373.1483E}.}
\label{Fig5}
\end{center}
\end{figure}

\begin{figure}
\begin{center}
\includegraphics[width=13.5cm,height=9cm]{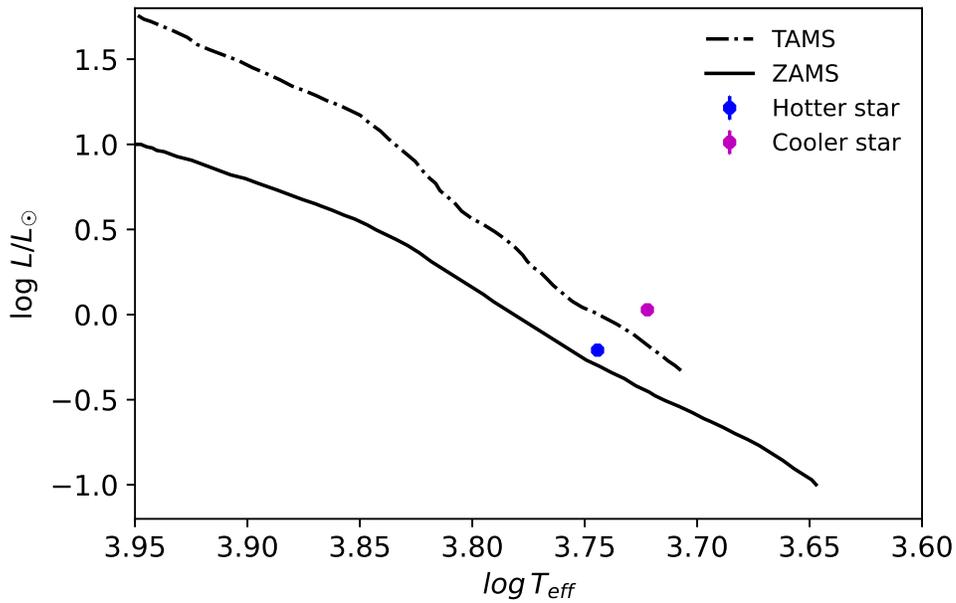}
    \caption{The hotter and cooler stars of the AH Mic system are shown on the H-R diagram.}
\label{Fig6}
\end{center}
\end{figure}

The H-R diagram depicted the evolutionary state of the components (Figure \ref{Fig6}). The cooler star (which is more massive) has already evolved away from the terminal-age main sequence (TAMS), and the hotter star stays in the main sequence belt and above the zero-age main sequence (ZAMS). According to the value of the mass ratio, the fill-out factor, inclination, and location of components on the H-R diagram, AH Mic follows the general pattern of the W-type W UMa systems.

\clearpage
\vspace{1.5cm}
\section*{Acknowledgments}
{This manuscript was prepared based on the Binary Systems of the South and North Project ($http://bsnp.info$). The goal of the project is to investigate contact binary systems in the northern and southern hemispheres using observatories from throughout the world.

We utilized the latest version of Gaia EDR3 (http://www.cosmos.esa.int/gaia) data from the European Space Agency (ESA) mission Gaia in this study. This work has made use of the SIMBAD database, operated by CDS, Strasbourg, France. Also, we used the latest version of the Phoebe related to this project website ($http://phoebe-project.org/$). PHOEBE is funded in part by the National Science Foundation (NSF 1517474, 1909109) and the National Aeronautics and Space Administration (NASA 17-ADAP17-68).}
We thank Fatemeh Davoudi for all her scientific assistance.

\vspace{1.5cm}
\section*{ORCID iDs}
\noindent Atila Poro: \url{https://orcid.org/0000-0002-0196-9732}\\
Mark G. Blackford: \url{https://orcid.org/0000-0003-0524-2204}\\
Selda Ranjbar Salehian: \url{https://orcid.org/0000-0002-5223-1332}\\
Esfandiar Jahangiri: \url{https://orcid.org/0000-0002-1576-798X}\\
Meysam Samiei Dastjerdi: \url{https://orcid.org/0000-0002-7502-3907}\\
Mohammadjavad Gozarandi: \url{https://orcid.org/0000-0002-5131-6801}\\
Reihaneh Karimi: \url{https://orcid.org/0000-0003-2295-0499}\\
Tabassom Madayen: \url{https://orcid.org/0000-0002-2653-3281}\\
Elnaz Bakhshi \url{https://orcid.org/0000-0002-8782-6045}\\
Farnad Hedayati: \url{https://orcid.org/0000-0003-0925-6806}\\

\clearpage
\bibliographystyle{aasjournal}
\bibliography{new.ms}

\end{document}